\documentclass[namedreferences]{SolarPhysics}
\usepackage[optionalrh,solaenum]{spr-sola-addons} 
\usepackage{graphicx}                    
\usepackage{color}                       
\usepackage{url}                         

\begin{document}

\begin{article}

\begin{opening}

\title{The Evolution of Sunspot Magnetic Fields Associated with a Solar Flare}

%
\author{Sophie~A.~\surname{Murray}$^{}$\sep
        D.~Shaun~\surname{Bloomfield}\sep
        Peter~T.~\surname{Gallagher}      
       }

%
\runningauthor{S.A.~Murray \emph{et al.}}
\runningtitle{Evolution of Sunspot Magnetic Fields Associated with a Solar Flare}


%
  \institute{$^{}$Astrophysics Research Group, School of Physics, Trinity College Dublin, Dublin 2, Ireland.\\
           S.~A.~Murray \\ email: \url{somurray@tcd.ie}\\  
            D.~S.~Bloomfield \\ email: \url{shaun.bloomfield@tcd.ie} \\
             P.~T.~Gallagher \\ email: \url{peter.gallagher@tcd.ie} \\
                     }

\begin{abstract}
Solar flares occur due to the sudden release of energy stored in active-region magnetic fields. To date, the pre-cursors to flaring are still not fully understood, although there is evidence that flaring is related to changes in the topology or complexity of an active region's magnetic field. Here, the evolution of the magnetic field in active region NOAA 10953 was examined using \textit{Hinode}/SOT-SP data, over a period of 12 hours leading up to and after a GOES B1.0 flare. A number of magnetic-field properties and low-order aspects of magnetic-field topology were extracted from two flux regions that exhibited increased Ca \textsc{ii} H emission during the flare. Pre-flare increases in vertical field strength, vertical current density, and inclination angle of $\approx$8$^\circ$  towards the vertical were observed in flux elements surrounding the primary sunspot. The vertical field strength and current density subsequently decreased in the post-flare state, with the inclination becoming more horizontal by $\approx$7$^\circ$. This behaviour of the field vector may provide a physical basis for future flare forecasting efforts.
\end{abstract}

%

\end{opening}

%
\section{Introduction}
 	\label{intro} 
 
Active regions in the solar atmosphere have complex magnetic fields that emerge from subsurface layers to form loops which extend into the corona. When active regions undergo external forcing, the system may destabilise and produce a solar flare, where energy stored in sunspot magnetic fields is suddenly released as energetic particles and radiation across the entire solar spectrum \cite{rust92,conlon08}. The initial impulsive phase of the flare is generally believed to be driven by magnetic reconnection, which leads to a change in the topology of the magnetic field, and energy stored in the field is released, accelerating coronal particles \cite{aschwanden05}.  The storage of magnetic energy in active regions is indicated by the degree of non-potentiality of sunspot magnetic fields \cite{regnierpriest07}. The processes leading up to reconnection and energy release are still not fully understood, and studying the links between solar flares and topology changes in active-region magnetic fields is an important step in understanding the pre-flare configuration and the process of energy release \cite{hewett08,conlon10}. 

Many early theoretical studies suggested a link between both the emergence of new flux and the shearing and twisting of field lines with the flare trigger mechanism (see \opencite{rust94}). Shearing is taken to mean that the field is aligned almost parallel to the neutral line rather than perpendicular to it, as would be observed in a potential field \cite{schmieder96}. \inlinecite{tanaka86} depicts a possible evolution of large-scale fields in a flare, with an ensemble of sheared fields containing large currents and a filament located above the neutral line in the pre-flare state. \inlinecite{canfield91} explored the importance of strong currents further, finding that sites of significant energetic-electron precipitation into the chromosphere were at the edges of regions of strong vertical current rather than within them. \inlinecite{metcalf94a} and \inlinecite{li97} subsequently found that flares do not necessarily coincide spatially with the locations of strong vertical current. More sophisticated flare models were later developed, \textit{e.g.} \inlinecite{antiochos98} described a ``breakout'' model for large eruptive flares, with newly-emerged, highly-sheared field held down by an overlying unsheared field. Field topology studies have been used to place constraints on theoretical models, \textit{e.g.} \inlinecite{mandrini06} reviewed a number of flaring active region topologies, finding that magnetic reconnection can occur in a greater variety of magnetic configurations than traditionally thought. The reader is referred to the reviews of \inlinecite{priestforbes02} and \inlinecite{Schrijver09}, and references therein, for more recent developments in eruptive event models.

Numerous observational studies have confirmed the importance of emergence and shearing to flare phenomena. \inlinecite{zirin93} investigated flux emergence and sunspot group motions, which resulted in complicated flow patterns leading to flaring. \inlinecite{wang94} used vector magnetograms to observe magnetic shear in five X-class solar flares; in all cases increasing along a substantial portion of the magnetic neutral line. They suggested flux emergence being key to eruption, but the increase in shear persisted much longer after the flare rather than decreasing as per model predictions. No definitive theoretical explanation was given. Recent evidence has furthered the idea that emerging-flux regions and magnetic helicity are crucial to the pre-flare state (\textit{e.g.} \opencite{liuzhang01}; \opencite{wang02}; \opencite{chandra09}), where magnetic helicity is a measure of magnetic topological complexity, \textit{e.g.} twists and kinks of field lines (see \opencite{canfield98}). Line-of-sight (LOS) magnetic-field observations have shown that photospheric fields change rapidly during large solar flares \cite{sudolharvey05,petriesudol09}. Other studies use improved extrapolation techniques to analyse the topology further, increasing our understanding of eruptions in the solar corona \cite{regnier06,georgoulis07}. Observing active-region magnetic fields around the time of flaring can be very beneficial, as magnetic-field properties have been found to be viable flare-forecasting tools \cite{gallagher02}. However, the LOS magnetic field alone cannot provide complete information on the changing magnetic field.

High spatial resolution observations of the solar magnetic-field vector can now provide more in-depth information on the true 3D topological complexities. In this paper we use spectropolarimetric measurements from the \textit{Hinode} spacecraft \cite{tsuneta08} to examine how sunspot magnetic fields evolve leading up to and after flare activity. In particular, differences in the magnetic-field vector between pre- and post-flare states are examined in the vicinity of a chromospheric flare brightening. Studying the evolution of the magnetic field before the flare with these improved observations could outline some new flare precursors that may of be use in flare forecasting, perhaps in terms of how soon a flare could be expected after certain conditions are met. Any changes observed after the flare compared to the pre-flare conditions should also give insight into how a flare might occur from this kind of region, testing the validity of currently proposed changes in magnetic topology during solar flares (\textit{e.g.} \opencite{pevtsovcanfieldzirin96}). In Section 2 we briefly discuss the observations and data analysis techniques used. Section 3 presents the main results, in particular the changes in vertical and horizontal field in Section 3.1, field orientation in Section 3.2, and derived low-order 3D magnetic properties in Section 3.3. Finally, our main conclusions and directions for future work are outlined in Section 4.

 \section{Observations and Data Analysis}
 	\label{observations} 

Active region NOAA 10953\footnote{\url{http://www.solarmonitor.org/region.php?date=20070426&region=10953}}~crossed the solar disk from 26 April 2007 to 9 May 2007. Previous studies of this region have found evidence of twisting, \textit{e.g.} \inlinecite{canou10} examined the magnetic structure of the region on 30 April 2007. Their reconstructed magnetic configurations exhibited twisted flux ropes along the southern part of the neutral line, similar to observations by \inlinecite{okamoto09} of twisted flux ropes emerging from below the photosphere. Here, we use observations of the main sunspot on 29 April 2007 recorded by the \textit{Solar Optical Telescope} (SOT: \opencite{suematsu08}) onboard \textit{Hinode}. Table~\ref{table} list the scan start and end times and pointing information. The simple structured active region comprised of a negative-polarity leading sunspot and opposite-polarity trailing plage, with an ``S-shaped'' filament visible over this time.  In addition, this region was the source of a low-magnitude GOES B1.0 solar flare: beginning at 10:34 UT; peaking at 10:37~UT; ending at 10:40~UT.

Four scans from the SOT spectropolarimeter (SP: \opencite{Kosugi07}) were used, with a scan duration of $\approx$32~minutes each. The temporal scan coverage was a critical reason for choosing this event, \textit{i.e.} three scans before the flare and one immediately after (Table~\ref{table}). Using multiple scans prior to the flare enables the non-flare related evolution of the magnetic-field properties to be analysed in detail, with changes over the flare able to be compared to this background evolution. No other flares occurred during the entire time period of observation, preventing the contamination of any of the scans.

SOT-SP recorded the Stokes $I$, $Q$, $U$, and $V$ profiles of the Fe \textsc{i} 6301.5~\AA~and 6302.5~\AA~lines simultaneously through a $0.16''\times164''$ slit. The Stokes spectral profiles were recorded with a spectral sampling of 21.5~m\AA, a field-of-view (FOV) of $164''\times164''$ (512 $\times$ 512 pixels), and an exposure time of 3.2~seconds per slit position (fast map mode). The raw SOT-SP data were calibrated using \textsf{sp$\_$prep.pro} from the \textit{Hinode}/SOT tree within the IDL \textsf{SolarSoft} library \cite{Freeland98}, which makes two passes through the data. The first determines the thermal shifts in the spectral dimension (in both offset and dispersion) across successive slit positions. The second pass corrects these thermal variations and merges the two orthogonal polarisation states.

The resulting Stokes $I$, $Q$, $U$, and $V$ profiles were inverted using the \textit{He-Line Information Extractor} (\textsf{H\textsc{e}LI\textsc{x}$^{+}$}: \opencite{Lagg04}) in order to derive the magnetic-field vector. \textsf{H\textsc{e}LI\textsc{x}$^{+}$} fits the observed Stokes profiles with synthetic ones obtained from an analytic solution of the Unno--Rachkovsky \cite{Unno56} equations in a Milne--Eddington atmosphere. The model atmosphere used in fitting the observed profiles comprised of one magnetic component with a local straylight component included. Optimal atmospheric parameters are obtained using \textsf{PIKAIA}, a genetic algorithm-based general purpose optimisation subroutine \cite{Charbonneau95}. A total polarisation threshold of $\approx$3$\times$10$^{-3}$~$I_c$ (\textit{i.e.} units of continuum intensity) was chosen, such that regions with values below this were not inverted.

\begin{table}[!t]
\caption{Summary of SOT-SP scan times on 29 April 2007.}
\begin{minipage}{8.7cm}
\begin{tabular}{ c c c c c }
\hline
  Scan Number & Begin Time 	& End Time 	& Centre of FOV \\
  			& (UT)			&  (UT)		&    (Solar X, Solar Y)           \\
  \hline
  1                        & 00:17     	& 00:49	   	&  -549$''$, -99$''$\\ 
  2                        & 03:30     	& 04:02	   	&  -525$''$, -98$''$\\ 
  3 			& 08:00     	& 08:32	  	&  -491$''$, -96$''$\\ 
Flare\footnote{Flare location corresponds to reconstructed RHESSI image peak.}	& 10:34	 	& 10:40		&   -476$''$, -150$''$\\ 
  4 			& 11:27	 	& 11:59	 	&  -464$''$, -95$''$\\
  \hline
  \label{table}
  \end{tabular}
  \vspace{-1.8\skip\footins}
   \renewcommand{\footnoterule}{}
\end{minipage}
\end{table}

The \textsf{AMBIG} routine \cite{Leka09}, which is an updated form of the Minimum Energy Algorithm \cite{Metcalf94}, was used to remove the  $180^{\circ}$ ambiguity in the LOS azimuthal angle. This procedure was chosen over other routines as it scored highly in the \inlinecite{Metcalfetal06} and \inlinecite{Lekaetal09} reviews on methods for resolving solar ambiguity angles. The routine simultaneously minimises the magnetic field divergence, $\nabla \cdot \mathbf{B}$, and vertical electric current density, $J_{z}$, for pixels above a certain noise threshold in transverse field strength. In this work we take a value of 150~G, whereby pixels with values below this level are determined using an iterative acute-angle-to-nearest-neighbors method \cite{Canfield93}. 

The resulting LOS inversion results were converted to the solar surface normal reference frame using the method outlined by \inlinecite{gary90}. The orthogonal magnetic-field components in the observers (\textit{i.e.} image, superscript ``i'') frame and solar surface normal (\textit{i.e.} heliographic, superscript ``h'') frame are related by,
\begin{eqnarray}
\label{gary90}
B_x^\mathrm{h} & = & a_{11}B_x\mathrm{^i}+a_{12}B_{y}^\mathrm{i}+a_{13}B_{z}^\mathrm{i} \ ,
\nonumber \\
B_{y}^\mathrm{h} & = & a_{21}B_{x}^\mathrm{i}+a_{22}B_{y}^\mathrm{i}+a_{23}B_{z}^\mathrm{i} \ ,
\nonumber \\
B_{z}^\mathrm{h} & = & a_{31}B_{x}^\mathrm{i}+a_{32}B_{y}^\mathrm{i}+a_{33}B_{z}^\mathrm{i} \ ,
\end{eqnarray}
where coefficients $a_{\mathrm{ij}}$ are defined in Equation (1) of \inlinecite{gary90}. In the image frame, $B_{z}^\mathrm{i}$ is the component along the LOS, and ($B_{x}^\mathrm{i}$, $B_{y}^\mathrm{i}$) define the plane of the image. In the heliographic frame, $B_{z}^\mathrm{h}$ is the component normal to the solar surface, and ($B_{x}^\mathrm{h}$, $B_{y}^\mathrm{h}$) lie in the plane tangent to the solar surface at the centre of the FOV. In terms of the field vector, $B_{x}^\mathrm{h} = \left | \mathbf{B} \right |\mathrm{sin}(\gamma)\mathrm{cos}(\phi)$, $B_{y}^\mathrm{h}=\left|\mathbf{B}\right|\mathrm{sin}(\gamma)\mathrm{sin}(\phi)$, and $B_{z}^\mathrm{h} = \left | \mathbf{B} \right |\mathrm{cos}(\gamma)$. Here, $\left | \mathbf{B} \right |$ is the absolute magnetic field strength, $\gamma$ is the inclination angle from the solar normal direction, and $\phi$ is the azimuthal angle in the ($B_{x}^\mathrm{h}$, $B_{y}^\mathrm{h}$) plane measured counter-clockwise from solar west.

\begin{figure} 
\centerline{\includegraphics[width=\textwidth,clip]{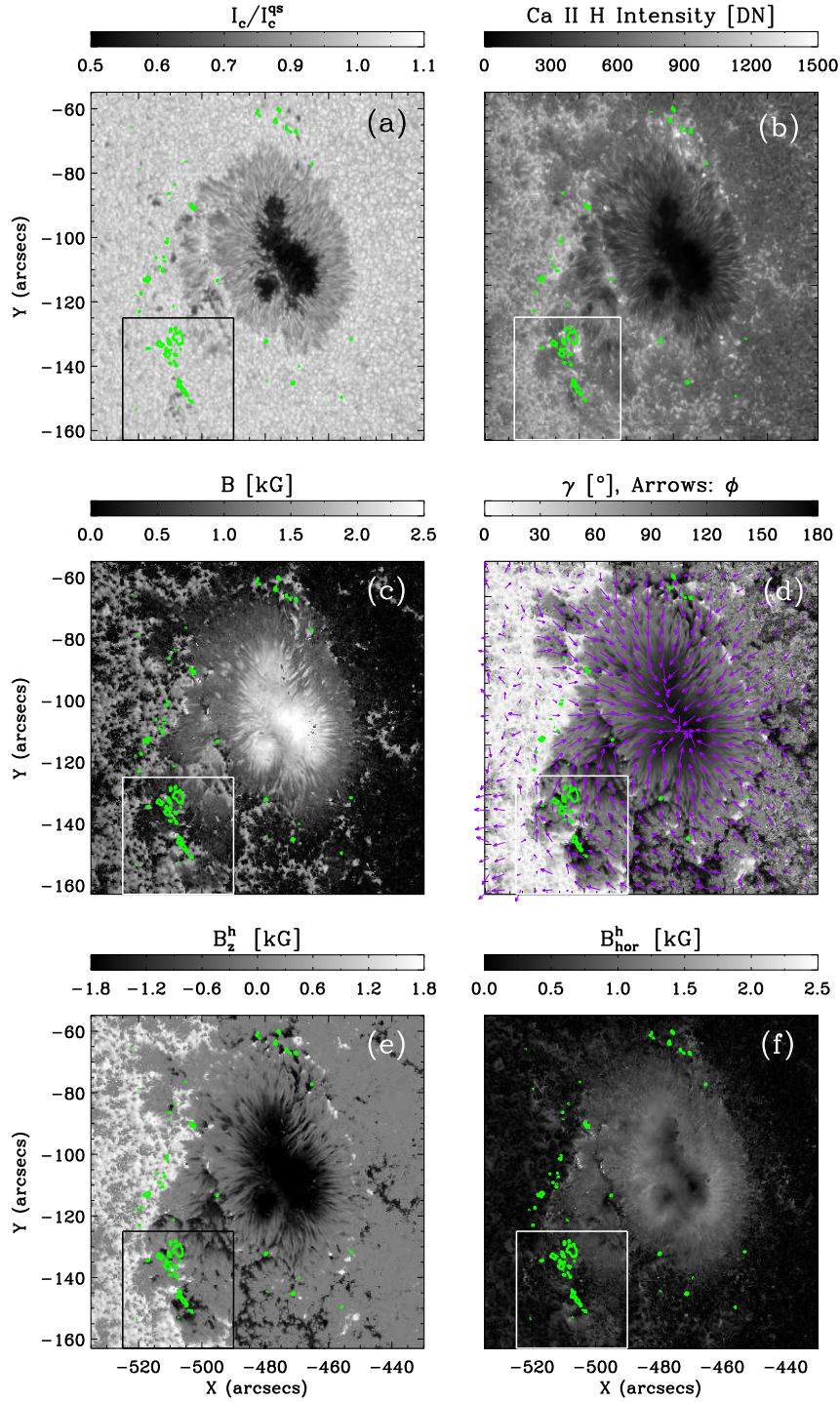}}
\caption{$108''\times108''$ FOV images showing the active region pre-flare state (08:00\,--\,08:32~UT): (a) continuum intensity; (b) Ca \textsc{ii} H intensity (08:16 UT); (c) absolute field strength; (d) inclination angle, with transverse magnetic field vectors overlaid as arrows (magenta); (e) vertical field strength; (f) horizontal field strength.  Green contours in all panels outline the significant Ca \textsc{ii} H flare brightening (at the 1300~DN level) observed at 10:37~UT. The sub-region selected for further analysis in Figure~\ref{figure_2} is indicated by the box in all panels. }
\label{figure_1}
\end{figure}

The scans were taken $\approx$three\,--\,four hours apart so it was necessary to correct for changes in scan pointing. To solve this, all scans were differentially rotated and their continuum intensity co-aligned to that of the third scan. Examples of observations from the third scan (\textit{i.e.} immediately preceding the flare) are shown in Figure~\ref{figure_1}, including \textit{Hinode}/SOT-SP continuum intensity (Figure~\ref{figure_1}a) and resulting magnetic field parameters from the \textsf{H\textsc{e}LI\textsc{x}$^{+}$} code after disambiguation and transformation to the solar normal reference frame: absolute magnetic-field strength (Figure~\ref{figure_1}c); inclination angle with azimuthal-angle vectors overlayed (Figure~\ref{figure_1}d); vertical field strength, $B_{z}^\mathrm{h}$ (Figure~\ref{figure_1}e); horizontal field strength, $B_\mathrm{hor}^\mathrm{h}=[(B_{x}^\mathrm{h})^2 +(B_{y}^\mathrm{h})^2]^{1/2}$ (Figure~\ref{figure_1}f). 

SOT \textit{Broadband Filter Imager} Ca \textsc{ii} H line images (3968~\AA) were also obtained close to the flare peak time, with a FOV of $108''\times108''$ (1024 $\times$ 1024~pixels$^2$). Figure~\ref{figure_1}b shows a Ca \textsc{ii} H image at the time of the third scan,  as well as contours of significant brightening at the time of the flare peak at 10:37~UT (1250~DN) overlaid on all other images. The brightening seems to be mostly located along the neutral line dividing the sunspot and plage regions in the East of the scan.  The location containing the most significant chromospheric flare brightening is found to the South East (SE) of the main sunspot, located near the trailing plage neutral line. A $35''\times40''$ box was chosen from this region for analysis. The sub-region was divided into two specific regions of interest, ROI 1 and ROI 2, defined by thresholding the signed field magnitude (\textit{i.e.} $|\mathbf{B}|$ times $-1$ or $+1$ for fields pointing in or out of the solar surface, respectively). ROI 1 was thresholded at -800~G and ROI 2 at -1000~G.  Both of these regions are small flux elements of the same polarity as the main sunspot, and are located SE of the main spot. They both lie close to the neutral line with the positive polarity plage (see Figure~\ref{figure_1}e). These two ROIs will be the focus of the magnetic field parameters studied.

 \section{Results}
 	\label{analysis} 
	
\begin{figure} 
\centerline{\includegraphics[width=1.0\textwidth,angle=180,clip]{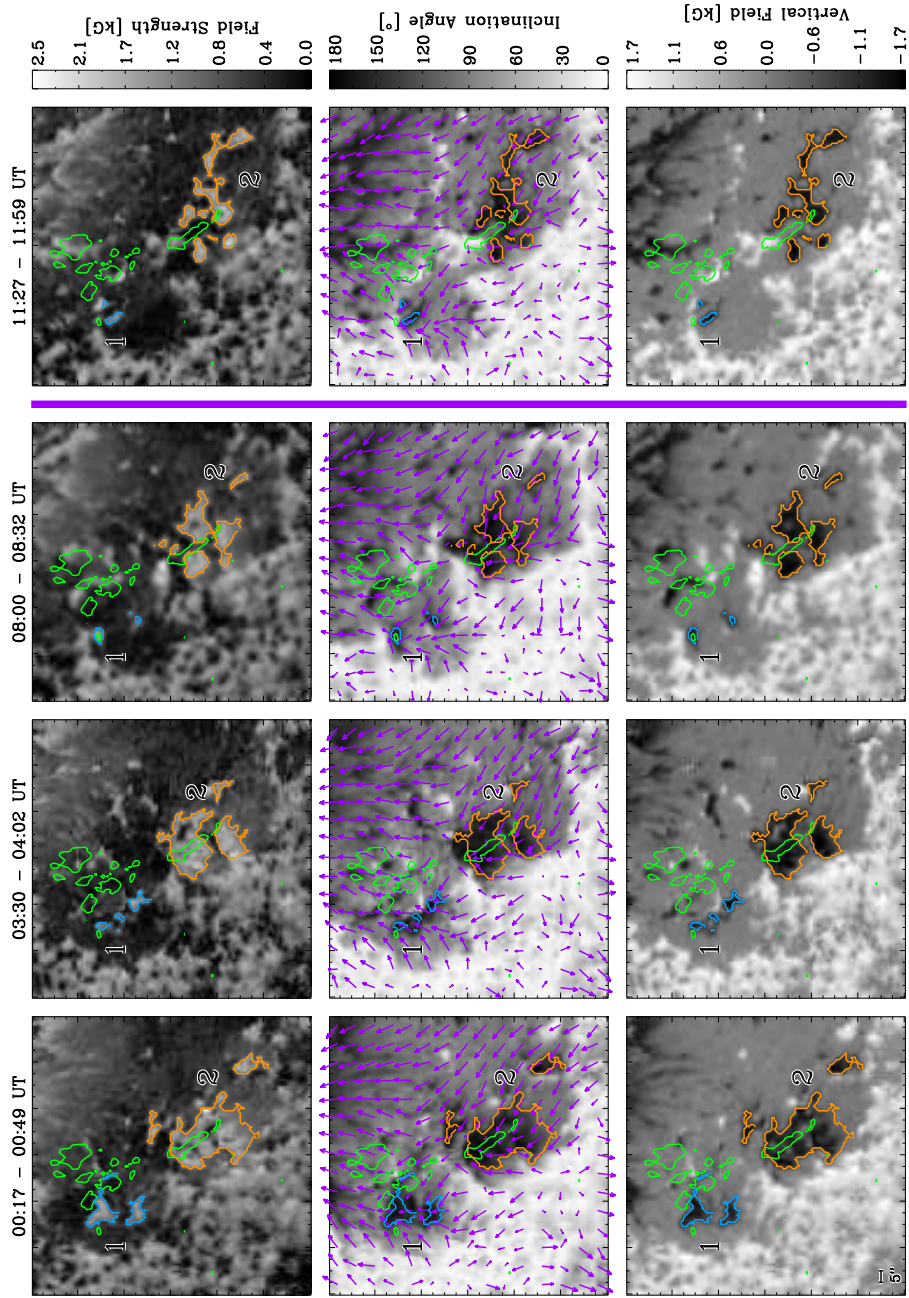}}
\caption{Evolution of the sub-region outlined in Figure~\ref{figure_1} at increasing scan times from left to right. Top to bottom: absolute magnetic field strength; inclination angle, with transverse magnetic field vectors overlaid as arrows (magenta); vertical field strength. Two regions of interest, ROI 1 and ROI 2, are numbered and defined by blue and orange contours, respectively. Green contours in all panels outline the significant Ca \textsc{ii} H flare brightening (at the 1300~DN level) observed at 10:37~UT, as per Figure~\ref{figure_1}. The time of flaring is indicated by a magenta vertical line between the third and fourth scans.}
\label{figure_2}
\end{figure}

Figure~\ref{figure_2} shows the temporal evolution of the magnetic field in the chosen sub-region over the four scans. ROI 1 fragments significantly from the first to the third pre-flare scans, and almost completely disappears after the flare. ROI 2 also fragments, but changes less than ROI 1. The chromospheric flare brightenings are located over and North West (NW) of ROI 1, and directly over ROI 2.

The parameters depicted in Figure~\ref{figure_2} were separately analysed in detail for both ROIs. The median and standard deviation of the values were extracted from all pixels within a ROI contour in each individual scan. Median values were used rather than other averaging methods due to their ease of interpretation and relative insensitivity to outlying values. The structure of the field was investigated in different ways: the vector field components (Section~\ref{vertical}); the field-orientation angles (Section~\ref{inclination}); signatures of magnetic non-potentiality (Section~\ref{divergence}).
 
\subsection{Vector Field Components}
\label{vertical}

\begin{figure} 
\centerline{\includegraphics[width=1.0\textwidth,clip]{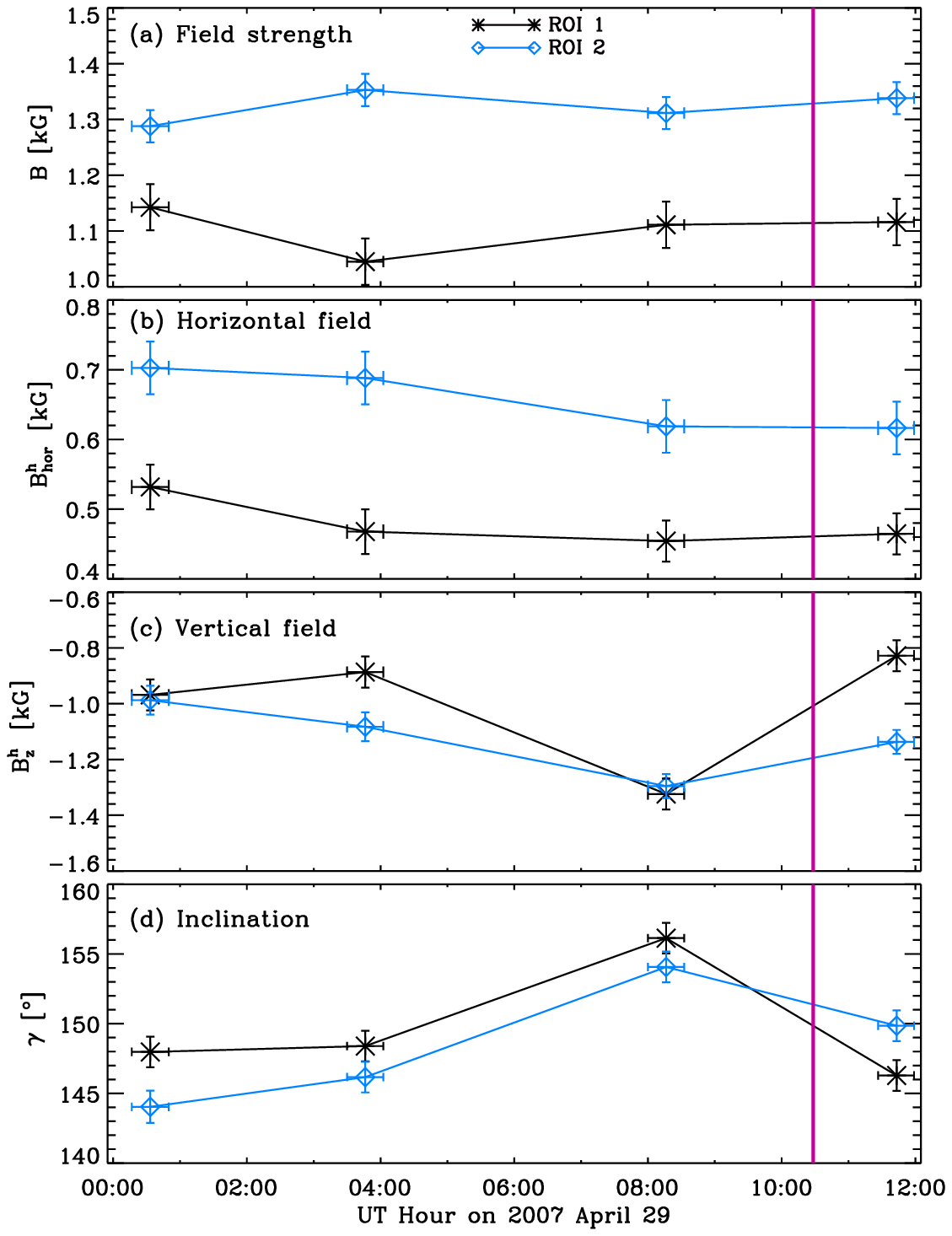}}
\caption{Temporal variation in the median values of: (a) absolute magnetic-field strength; (b) horizontal field strength; (c) vertical field strength; (d) inclination angle. Values for ROI 1 are plotted with black asterisks, and ROI 2 with blue diamonds. Vertical bars indicate the standard deviation of the property within the ROI, while horizontal bars delimit the scan duration. The vertical line between the third and fourth scans marks the flare peak time. With values of inclination being beyond $90^{\circ}$, increasing values indicate the field becoming more vertical.}
\label{figure_3}
\end{figure}

Changes in ROI median values of the field magnitude, vertical field, and horizontal field were calculated in each scan (\textit{i.e.} values from all pixels in the thresholded contours of a ROI). Figure~\ref{figure_3} depicts time lines of these ROI median values, with vertical bars representing the ROI standard deviation and horizontal bars depicting the scan duration. The magnetic field strength in Figure~\ref{figure_3}a varies little over all the scans within 1-$\sigma$ errors, with only a slight decrease in the second scan for ROI 1. The horizontal field strength, given in Figure~\ref{figure_3}b, shows only a slightly decreasing trend over the scans. The main source of interest here comes from the vertical field strength.

The vertical field median value also marginally changes within the spread of ROI values between the first two scans, as can be seen in Figure~\ref{figure_3}c. However, substantial variations are found between both the second and the third scans, as well as the third and fourth scans. An increase in vertical field magnitude is found between the second and third scans, increasing by $\approx$440~G for ROI 1 and $\approx$210~G for ROI 2. After the flare (\textit{i.e.} some time between the third and fourth scans) $B_{z}^\mathrm{h}$ decreases by $\approx$500~G for ROI 1 and $\approx$160~G for ROI 2. It is likely that the changes prior to the flare are linked to the energy storage mechanism in the ROIs, while the changes over the course of the flare are due to the energy release. However, it is unclear from the median field magnitude measurements how the field structure is changing before and after the flare. Thus, field orientation was investigated further.

\subsection{Field Orientation}
\label{inclination}
The median inclination angle was also extracted from both ROIs and is included in Figure~\ref{figure_3}d. A similar trend in inclination evolution is seen to the vertical-field evolution. Again no changes of significance are found between the first two scans, with large changes observed between the second and third scans and after the flare. An increase in inclination is found in the third scan, with field becoming more vertical by  $\approx$8$^{\circ}$ for both ROI 1 and ROI 2. After the flare, inclination decreases (\textit{i.e.} becomes more horizontal) by $\approx$10$^{\circ}$ for ROI 1 and $\approx$4$^{\circ}$ for ROI 2. These results support the idea that the field in both ROIs becomes more vertical $\approx$6.5\,--\,2.5 hours before the GOES B1.0 flare and more horizontal within $\approx$one hour after the flare has ended. It is interesting to note that the location of the field change is near the neutral line with the plage region, in a negative polarity region to the SE of the sunspot.
 
To put the changes in field parameters observed over the scans into context, it is worth estimating where the field lines in ROI 1 and ROI 2 are connected to by examining the direction of the transverse magnetic field vectors (overlaid on the inclination scans in Figure~\ref{figure_2}). However, the true connectivity cannot be determined from 2D results and the necessary 3D extrapolations of the region will be investigated in a future paper. As a first guess towards the possible connectivity, the field in ROI 1 seems to be generally pointing towards a northerly direction in Scan 1 and Scan 2, becoming increasingly more NE in Scan 3 and Scan 4. In ROI 2, the field is pointing in a general NE direction in the first scan, pointing in an increasingly more easterly direction as time progresses, finally becoming more NE after the flare. It seems that the plage region SE of the ROIs extends towards the NW (\textit{i.e} between the ROIs) as the scans progress, before pinching off after the flare. It is difficult to determine by eye exactly where the field may be connected to over the scans, especially if relying on median values of small groups of pixels. We surmise a region of plage NE of ROI 1 to be a likely connection point. The fourth scan in Figure~\ref{figure_2} also indicates a possible connection between ROI 2 and the portion of intersecting plage that first extends between the ROIs before ``pinching off'' after the flare. Studying the field distributions within the ROIs is necessary to fully understand the evolution.

\subsection{Signatures of Non-Potentiality}
\label{divergence}
\begin{figure}[!t]
\centerline{\includegraphics[width=1.0\textwidth,clip]{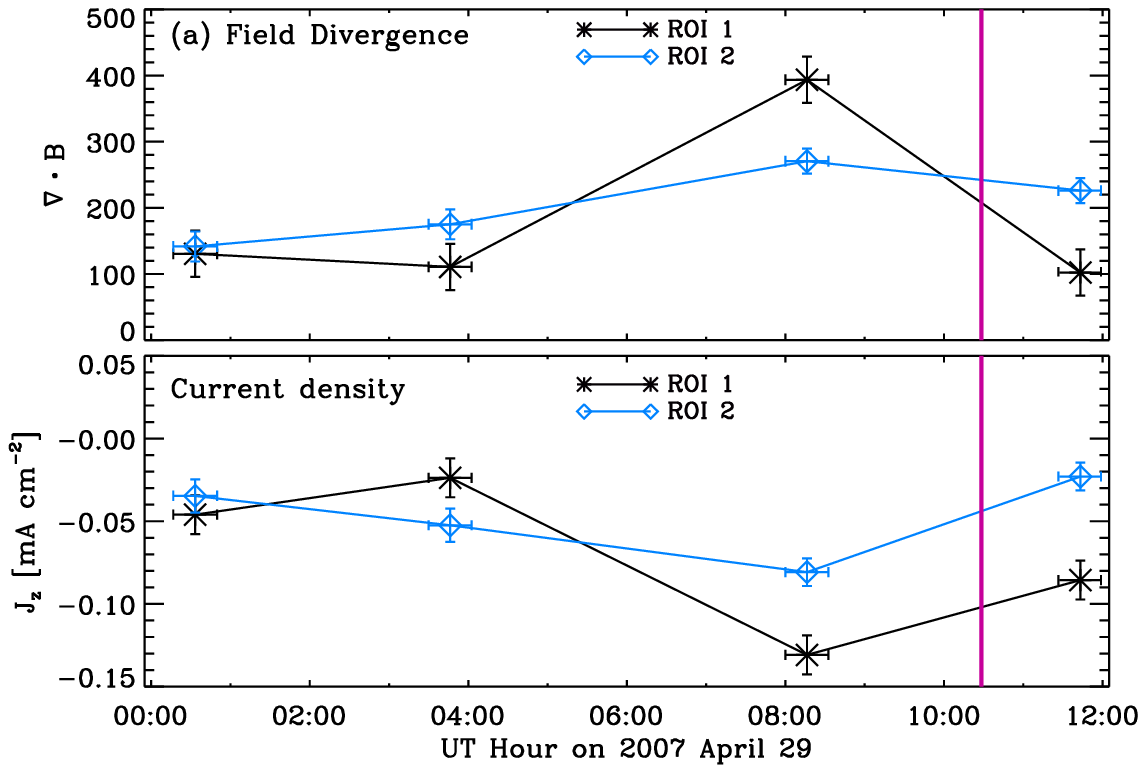}}
\caption{Temporal variation in the median values of vertical current density. Values for ROI 1 are plotted with black asterisks, and ROI 2 with blue diamonds. Vertical bars indicate the standard deviation of the property within the ROI, while horizontal bars delimit the scan duration. The vertical line between the third and fourth scans marks the flare peak time.}
\label{figure_4}
\end{figure}

The vertical current density was calculated by the method of \inlinecite{crouch08}, as implemented in the \textsf{AMBIG} code. Median values of all pixels within the contours for each ROI are presented in Figure~\ref{figure_4}, with vertical bars again showing the ROI standard deviation. A familiar trend is seen between the first and second scans (\textit{i.e.} no change within the spread of values in either ROI). Negative vertical current density increases in magnitude in the pre-flare state from the second to third scans by $\approx$0.11~$\mathrm{mA} \, \mathrm{cm^{-2}}$ for ROI 1 and by $\approx$0.03~$\mathrm{mA} \, \mathrm{cm^{-2}}$ for ROI 2. The magnitude subsequently decreases by $\approx$0.07~$ \, \mathrm{mA} \, \mathrm{cm^{-2}}$ in both ROI 1 and ROI 2. Changes in ROI 1 parameters are much more distinct than in ROI 2, as was also seen in field inclination and vertical field strength. Thus, stronger currents appear in both regions before the flare occurs, dropping back to earlier background values after the flare. An increase in current density before the flare indicates an emergence or build-up of non-potentiality in the field, with these observed changes likely to be linked to energy build-up before the flare, and energy release during to the flare.

 \section{Discussion and Conclusions}
 	\label{discussion} 

An $\approx$8$^{\circ}$ change in field inclination towards the vertical was found leading up to the flare,  with a $\approx$7$^{\circ}$ return towards the horizontal afterwards. Note that the inclination changes towards the vertical had occurred by $\approx$2.5 hours before the flare onset, with no changes observed $\approx$6.5\,--\,10 hours beforehand.  \inlinecite{Schrijver07} states that the energy build-up phase can last for as much as a day in an active region, so it is interesting to see such short time-scale changes. Previous studies have also reported changes in field orientation after a flare. \inlinecite{li09} found an inclination angle change of $\approx$5$^{\circ}$ towards the horizontal in a region of enhanced G-band intensity after an X-class flare, and the inclination becoming more vertical by $\approx$3$^{\circ}$ in a region of diminished G-band intensity.  Although their study focuses on penumbral regions, the region becoming more horizontal after the flare is located close to the flaring neutral line, similar to our findings. This concept is also mentioned in some theoretical studies, \textit{e.g.} \inlinecite{hudson08} predicted that the photospheric magnetic fields close to the neutral line would become more horizontal in a simple flare restructuring model.

Examining previous findings of transverse field changes, \inlinecite{wang02} used vector magnetogram observations to find an impulsive increase of the transverse-field strength and magnetic shear after three X-class flares. \inlinecite{li09} also found a transverse field increase of 20\% after an X3.4 flare. We found no significant changes in transverse-field strength either immediately before or after the flare. However, the difference in active regions must be noted, with the \citeauthor{wang02} and \citeauthor{li09} works focusing on higher-magnitude flares from $\delta$ sunspot groups. Our insignificant changes in the transverse-field strength are explained by the competing field strength and inclination changes before and after the flare. For example, a large increase in inclination angle for ROI 1 (Figure~\ref{figure_3}d) between the second and third scans is accompanied by a slight increase in field strength (Figure~\ref{figure_3}a), giving approximately no change in the horizontal field (Figure~\ref{figure_3}b). The \inlinecite{li09} result supports the reconnection picture of \inlinecite{liu05}, whereby newly connected fields near the magnetic neutral line contributed to field inclination becoming more horizontal. This picture suggests that the field lines after the flare in our study become newly-reconnected, low-lying, more horizontal field lines near the flaring neutral line. 

Vertical-field magnitude was found in our results to increase in both ROIs before the flare, and decrease by approximately the same amount afterwards. \inlinecite{wang02} examined LOS magnetograms as well as vector data, finding an increase in magnetic flux of the leading polarity in six X-class flares. \inlinecite{sudolharvey05} used longitudinal magnetogram data from the \textit{Global Oscillation Network Group} to find abrupt and permanent changes in the LOS magnetic field after 15 X-class flares. They found decreases in vertical field twice as often as increases: in 75\% of cases the magnetic-field change occurred in less than ten minutes. \citeauthor{sudolharvey05} quote median LOS field changes of 90 G and found that the strongest field changes typically occur in penumbrae. This behaviour of decreasing vertical field is reflected in our findings, although we observe larger changes of $\approx$330~G in a region outside of the penumbra.

Our observations find an increase in negative vertical current density within $\approx$6.5\,--\,2.5 hours before the flare, with a decrease towards the initial pre-flare values after the flare. Strong emerging currents have often been linked with flare triggers, \textit{e.g.} \inlinecite{su09} observed the current density for the same active region three days later when a C8.5 flare occurred, finding strong currents along the field lines. \inlinecite{canou10} also examined the vertical current density for the same active region using a different extrapolation method than \inlinecite{su09}. The extrapolation found footpoints of the twisted flux ropes to be anchored in a region of significant vertical current (\textit{i.e.} in the core of the flux region rather than along the field lines). They observed the breakdown of the force-free assumption along the neutral line due to non-zero vertical current density and suggested that this could be due to the emergence of the twisted flux ropes, or perhaps the presence of non-null magnetic forces. They also determined that enough free magnetic energy existed to power the C8.5 flare studied by \inlinecite{su09} and a C4.2 flare a few days later. Similar mechanisms could possibly be at work to cause the earlier lower-magnitude flare examined in this paper. 

\inlinecite{regnierpriest07} noted the discrepancies that exist between using different extrapolation methods. They found that strong currents present in the magnetic configuration were responsible for highly twisted and sheared field lines in a decaying active region. In contrast, weak currents existed in a newly emerged active region. They also suggest a strong dependance of vertical current density on the nature of the active region, \textit{e.g.} the stage of the regions' evolution or the distribution of the sources of magnetic field.  Most previous work has focused on considerably more complex active regions that produce M- or X-class flares, so it is important to note that distinct changes in the magnetic field were still observed for this B-class flare.
 
It is worth mentioning that \inlinecite{okamoto09} observed converging motions in Ca \textsc{ii} H movies of the same active region as this paper, which they describe as driven by moat flows from the sunspot towards the trailing plage neutral line (\textit{i.e.} near our ROI locations).  \inlinecite{Schrijver00} mention typical spatial scales of moat flow regions of  $\approx$10 - 20~Mm measured from the outer edge of the penumbra. This suggests our ROIs lie on the outer edge of the moat flow region, and perhaps a moving magnetic feature was being driven towards the plage region. This driving would cause the field near the neutral line to become more vertical before the flare, as per our results, and might explain the pre-flare energy build-up phase. The field would then relax and become more horizontal after the energy release, as we found. The driving could be related to converging motions towards the neutral line highlighted in a number of MHD simulations (\textit{e.g.}  \citeauthor{amari03} 2003). \citeauthor{amari03} mention a three part magnetic structure associated with their model's disruption phase, with a twisted flux rope running through a global arcade and above small loops. These newly formed small loops, described as due to reconnection, are perhaps indicative of the more horizontally inclined post-flare field of this paper compared to pre-flare build up values.

Further work is planned to clarify the connectivity of the two regions of interest and changes in the 3D topology. Our resulting disambiguated field vector will be used as an input to a magnetic field extrapolation to determine various topology measures (\textit{e.g.} numbers and locations of nulls, separatrix layers). However, it is interesting to see such clear changes in field vector characteristics (such as inclination, magnetic divergence, and vertical current density) leading up to and after the flare, before making high order calculations of 3D topology. These forms of field orientation changes could prove to be useful precursors for flare forecasting in the future. 

%

%
 \begin{acks}
\textit{Hinode} is a Japanese mission developed and launched by ISAS/JAXA, with NAOJ as domestic partner and NASA and STFC (UK) as international partners. It is operated by these agencies in co-operation with ESA and NSC (Norway). S.A.M. is supported by the AXA Research Fund while D.S.B. is supported by the European Community (FP7) under a Marie Curie Intra-European Fellowship for Career Development.  
 \end{acks}

%
%
 \bibliographystyle{spr-mp-sola-cnd} 
 \bibliography{bibliography}  
%

\end{article} 
\end{document}